# Lévy-Kolmogorov scaling of turbulence

W. Chen and H.B. Zhou

*Institute of Applied Physics and Computational Mathematics, P.O. Box 8009, Division Box 26,, Beijing 100088, China (chen_wen@iapcm.ac.cn)*

**The Kolmogorov scaling law[1,2] of turbulences has been considered the most important theoretical breakthrough in the last century. It is an essential approach to analyze turbulence data present in meteorological, physical, chemical, biological and mechanical phenomena[3,4]. One of its very fundamental assumptions is that turbulence is a stochastic Gaussian process in small scales[5]. However, experiment data at finite Reynolds numbers have observed a clear departure from the Gaussian[5-9]. In this study, by replacing the standard Laplacian representation of dissipation in the Navier-Stokes (NS) equation with the fractional Laplacian[10,11], we obtain the fractional NS equation underlying the Lévy stable distribution which exhibits a non-Gaussian heavy trail and fractional frequency power law dissipation[12]. The dimensional analysis of this equation turns out a new scaling of turbulences, called the Lévy-Kolmogorov scaling, whose scaling exponent ranges from -3 to -5/3 corresponding to different Lévy processes and reduces to the limiting Kolmogorov scaling -5/3 underlying a Gaussian process. The truncated Lévy process and multi-scaling due to the boundary effect is also discussed. Finally, we further extend our model to reflecting the history-dependent fractional Brownian motion.**

Turbulence occurs all over nature from the atmosphere to the oceans to electronics to inside stars and internal combustion chambers. Scaling methods are used to explore hidden structures in the random behavior of turbulent fluid flow even without a detailed solution of the equations of motion. In the limit of vanishing viscosity (i.e., infinite Reynolds number), Kolmogorov's celebrated scaling of turbulence $E(k) = C\varepsilon^{\frac{2}{3}}k^{-\frac{5}{3}}$ is established[1,2], where $E(k)$ is the energy spectrum, $C$ denotes a absolute constant, and $\varepsilon$ represents the kinetic energy dissipation rate and is considered scale-independent[13]. In essence, the Kolmogorov -5/3 scaling characterizes the statistical similarity of turbulent motion at small scales based on the argument of local homogeneous isotropy [14]. To some extent, the scaling law has been validated by numerous experimental and numerical data of sufficiently high Reynolds number turbulence[3,4,15]. However, recent experiments[6,7] using high speed optical techniques reveal that the statistics of the Lagrangian acceleration manifests distribution profiles with long heavy tails, indicative of strong non-Gaussian process. This contradicts the very fundamental foundation of the Kolmogorov theory that turbulence obeys Gaussian distribution[5]. Therefore, the -5/3 scaling does not fit real-world turbulences at finite Reynolds numbers.

It has be suggested that the Lévy $\beta$-stable distribution is a proper statistical approach to accommodate heavy tails widely observed in the probability density function (PDF) of turbulence quantities[5,16], where $\beta$ is the Lévy stability index and ranges from 0 to 2. And the Gaussian distribution is its limiting $\beta=2$ case[17]. On the other hand, the fractional Laplacian is a non-local (integro-differential) and positive definite operator[11,18] underlying the Lévy process in a variety of physical master equations such as the Fokker-Planck equation[17] and the anomalous diffusion equation[18-20]. By replacing the Laplacian representation of dissipation process with the fractional Laplacian in the standard NS equation, this study introduces a fractional NS equation

$$\frac{\partial \mathbf{u}}{\partial t} + \mathbf{u} \cdot \nabla \mathbf{u} = -\frac{1}{\rho}\nabla p - \frac{1}{\widetilde{\mathrm{Re}}}\left(-\Delta\right)^{\beta/2}\mathbf{u}, \quad \beta \in (0,2], \tag{1}$$

where $\mathbf{u}$ represents velocity vector, $\widetilde{\mathrm{Re}}$ is the scaled Reynolds number, and the fractional Laplacian $\left(-\Delta\right)^{\beta/2}$ serves as a stochastic driver and guarantees the positive definiteness of energy dissipation. The essence of fractional NS equation (1) is that the constitutive equation in turbulence may not obey the classical Newtonian gradient law.

Using a Kolmogorov-like argument, the dimensional analysis of equation (1) leads to the energy spectrum

$$E(k) = \widetilde{C}\widetilde{\varepsilon}^{\frac{2}{3}}k^{-\frac{9-2\beta}{3}}, \tag{2}$$

where $\widetilde{C}$ and $\widetilde{\varepsilon}$ are respectively the scaled $C$ and $\varepsilon$ by parameter $\beta$. The corresponding exponent of decay power law ranges from -3 to -5/3, in which the upper limiting $\beta=2$ leads to the Kolmogorov -5/3 scaling corresponding to the Gaussian distribution and the standard NS equation, while the lower limiting $\beta \rightarrow 0$ coincides with -3 power scaling which has been observed in experiments[21]. Mandelbrot[22] pointed out that the intermittent (non-Gaussian) property of turbulence calls for a power law of energy spectrum having exponent -5/3-$c$ ($c \geq 0$). But he did not quantitatively clarify his correction as this study did. Since the scaling (2) generalizes the Kolmogorov scaling and underlies Lévy statistics, we call it the Lévy–Kolmogorov scaling, which reflects anomalous transportation of kinetic energy, as evidenced in plasma turbulence[23]. For example, the turbulent fluids having viscosity $\beta=2$ correspond to the Kolmogorov scaling, but the boundary layer turbulence is known to have $\beta=1/2$, a strong non-Gaussian Lévy process[12], leading to -8/3 scaling.

Unlike Gaussian process, Lévy process does not have finite moments of second or higher order. And the truncated Lévy distribution was thus proposed in turbulence modeling[24], in which the long fat tails of algebraic decay of the original Lévy distribution is truncated and replaced by the corresponding Gaussian distribution of exponential decay, and then the divergent second moments are cured. However, this truncation is somewhat arbitrary and the truncated Lévy distribution can no longer underlie the fractional Laplacian in the governing equation. It is noted that the standard Lévy distribution and fractional Laplacian are defined under infinite domain. However, the real-world turbulences all have finite Reynolds numbers, namely, the finite size of turbulence region. From this view, we have to take account of boundary effect into consideration. The standard definition of the fractional Laplacian under infinite domain encounters hypersingularity[10], which corresponds to the infinite moment of the second and higher orders of Lévy distribution[19,20]. Chen[11] recently introduced a new definition of fractional Laplacian under finite domain (equation 17 in ref. 11) which naturally includes boundary conditions and eliminates hypersingularity. Accordingly, the Lévy distribution corresponding to the fractional Laplacian of finite domain in the NS equation (1) is truncated in terms of boundary conditions and has the finite square moment. It is known that the truncated Lévy distribution gives rise to the intermittency (finite scaling range) and multifractality (multi-scaling) of turbulence phenomena, which has long been observed as a rule rather than exception in turbulence[8,9]. Multifractality suggests that the Lévy-Kolmogorov scaling exponent is not universal over all scales. It is noted that the standard dimensional analysis of the NS equation leading to the scaling power law does not consider the effect of boundary conditions.

Turbulence is observed experimentally and numerically to have anomalous diffusion[23,25,26]. The above fractional NS equation (1) can be reduced to the well known anomalous diffusion equation underlying a time-dependent $\beta$-stable Lévy process ($\eta = 2/\beta$, $0<\beta<2$)[16,17]. Alternatively, by replacing the fractional Laplacian dissipation in equation (1) with a mixed operator of the fractional time derivative and the Laplacian $-\dfrac{1}{\mathrm{Re}}\dfrac{\partial^{\mu}}{\partial t^{\mu}}\Delta\mathbf{u}$, $0\leq\mu<1$, we have a new fractional NS equation underlying the fractional Brownian motion. The dimensional analysis of this equation leads to a scaling,

$$E(k) = \hat{C}\hat{\varepsilon}^{\frac{2}{3}}k^{-\frac{5-3\mu}{3-\mu}} , \qquad (3)$$

whose lower and upper limiting exponents are the Kolmogorov scaling -5/3 and -1, respectively. Furthermore, the combination of the fractional Laplacian and the fractional time derivative can be used in the NS equation to represent anomalous dissipation underlying the statistic paradigms of the fractional Brownian and the Lévy processes,

$$\frac{\partial \mathbf{u}}{\partial t} + \mathbf{u} \cdot \nabla \mathbf{u} = -\frac{1}{\rho}\nabla p - \frac{1}{\mathrm{Re}}\frac{\partial^{\mu}}{\partial t^{\mu}}(-\Delta)^{\beta/2}\mathbf{u} , \qquad (4)$$

where $\beta \in (0,2]$ when $\mu$=0; $\mu \in [0,1)$ when $\beta$=2; and *Re* represents the time-space scaled Reynolds number. The above NS equation (4) reflects spatial and temporal fractal irregularity and memory effect of turbulent motions. The energy transfer scaling of turbulence is thus split into three phases: sub-transport (subdiffusion, fractional Brownian motion) before Kolmogorov scaling range, normal-transport (Kolmogorov scaling), and super-transport (Lévy-Kolmogorov scaling, e.g., ballistic transport with $\beta$=1, $\mu$=0 and -7/3 scaling). The orders $\beta$ and $\mu$ of the fractional derivative are considered time and space fractal dimensions[22]. Serving as an example, we consider the elastic turbulence of polymer solutions[27], whose equation for motion differs from the standard NS equation and has to reflect the history-dependent motion. The above NS equation (4) reflects the constitutive relationship between stress and the fractional time derivative representation of the deformation rate in a polymer flow and is a competitive alternative[11,12] to the conventional nonlinear constitutive model[27]. For the case *β=2* and *μ≠0,* the scaling exponent of polymer fluid turbulence is not the Kolmogorov index -5/3 even in the limit of infinite Reynolds number.

For a further analysis, the NS equation (4) is reduced to the well-known anomalous diffusion equation[18,19]

$$\frac{\partial^{1-\mu}\mathbf{u}}{\partial t^{1-\mu}} + \gamma(-\Delta)^{\beta}\mathbf{u} = 0, \tag{5}$$

whose corresponding discrete Lagrangian stochastic model is defined by the time evolution of the mean square displacement of diffusing particle movements $\langle \Delta x^2 \rangle \propto \Delta t^{\eta}$, where $\eta=2(1-\mu)/\beta$, $\Delta x$ represents distance, $\Delta t$ denotes time interval, the brackets represent the mean value of random variables (e.g., a collection of particles). Accordingly we can compare our theoretical predictions with experimental data, where the motion of tracer particles in turbulent flows is measured. For instance, the well-known $\langle \Delta x^2 \rangle \propto \Delta t^3$ in turbulence is first obtained by Richardson[28] to explain experimental measurements and results in $\mu=0$, $\beta=2/3$ in the NS equation (4) and -23/9 scaling. It is stressed that the corresponding anomalous diffusion equation (5) of fractional time-space derivatives is physically more reasonable than the Richardson's diffusion equation with a space- and time-dependent diffusion coefficient for deriving $\langle \Delta x^2 \rangle \propto \Delta t^3$, since integer-order differentiability in the latter may not exist in turbulent velocity flow fields.

It is known that Gaussian process corresponds to the normal diffusion ($\eta=1$), while Lévy process reflects the superdiffusion (long-range spatial correlation, $\eta \succ 1$) and the fractional Brownian motion underlies the subdiffusion ($\eta \prec 1$) which manifests history-dependent (long-range temporal correlation) motion. Once the signature $\eta$ of anomalous diffusion is known in turbulence, it is straightforward to derive the order of fractional derivatives of our NS equation model and then the scaling exponent. For a subdiffusion case, $\langle \Delta x^2 \rangle \propto \Delta t^{0.5}$ is found in magnetic field turbulence and cosmic-ray transport in the interstellar medium[23,26,29]. By using the previous formulas, it is simple to find $\mu=0.5$, $\beta=2$ in NS equation (4) and the corresponding scaling exponent -7/5.

In summary, this study introduced the fractional Laplacian NS equation and then presented a new Lévy-Kolmogorov -3 to -5/3 scaling of turbulence. The fractional time derivative was also used in the representation of dissipation in the NS equation underlying the history-dependent fractional Brownian motion and leads to the scaling law -5/3 to -1. Warhaft[30] pointed out "Apart from noting the presence of non-Gaussian tails, no deeper analysis of the shape of the *pdf*s has been made. Because the connection of these models to the Navier-Stokes equations is tenuous,...". In this study, the first attempt was made to explicitly connect non-Gaussian statistics of turbulence and the NS equation, where the fractional derivative representations reflect the strong influence of the viscose property of fluids on turbulence and vice versa. It is also worth noting that the direct numerical simulation of the fractional NS equation will be more challenging than that of the standard NS equation, since the fractional derivatives are a non-local operator and will result in the full matrix of numerical discretization which is usually computationally very costly.

**References**

1. Kolmogorov, A. N., Dissipation of energy in the locally isotropic turbulence. *Dok. Akad. Nauk SSSR* **31**, 538-540 (1941).
2. Obukhov, A. M., Structure of the temperature field in turbulent flow, *Isv. Akad. Nauk SSSR, Ser. Geogr. Geophys.*, **13**, 58–69 (1949).
3. Monin A.S. and Yaglom A.M. *Statistical Fluid Mechanics: Mechanics of Turbulence Volume 1* (Cambridge: The MIT Press, 1971).
4. Monin A.S. and Yaglom A.M. *Statistical Fluid Mechanics: Mechanics of Turbulence Volume 2* (Cambridge: The MIT Press, 1975).
5. Shlesinger, M.F., West, B.J. and Klafter, J., Levy dynamics of enhanced diffusion: Application to turbulence, *Phys. Rev. Lett.*, **58**(11), 1100-1103, 1987.
6. Voth, G. A., Satyanarayan, K. and Bodenschatz, E., Lagrangian acceleration measurements at. G/A, large Reynolds numbers, *Phys. Fluids*, **10**, 2268 (1998).
7. Porta, A. L. Voth, G.A., Crawford, A.M., Alexander, J., Bodenschatz, E., Fluid particle accelerations in fully developed turbulence, *Nature*, **409**, 1017-1019 (2001).
8. Sreenivasan K.R. and Antonia R.A. The phenomenology of small-scale turbulence. *Annual Reviews of Fluid Mechanics*, **29**, 435-472 (1997).
9. Vedula, P. & Yeung, P. K. Similarly scaling of acceleration and pressure statistics in numerical simulations of isotropic turbulence, *Phys. Fluids* **11**, 1208-1220 (1999).
10. Samko, S. G., Kilbas, A. A., & Marichev, O. I., *Fractional Integrals and Derivives: Theory and Applications* (Gordon and Breach Science Publishers, London, 1993).

11. Chen, W. and Holm, S., Fractional Laplacian time-space models for linear and nonlinear lossy media exhibiting arbitrary frequency power law dependency, *J. Acoust. Soc. Am.* **115**(4), 1424-1430 (2004).

12. Blackstock, D. T., Generalized burgers equation for plane waves, *J. Acoust. Sco. Am.* **77**(6), 2050-2053 (1985).

13. Shraiman, B. I. and Siggia, E. D. Scalar turbulence. *Nature* **405**, 639-646 (2000).

14. Lesieur M. *Turbulence in Fluid*. the Netherlands: Kluwer Academic Publishers,. 1997.

15. Batchelor G.K. *The Theory of Homogeneous Turbulence* (Cambridge, Cambridge University Press, 1953).

16. Takayasu, H., Stable distribution and levy process in fractal turbulence. *Progr. Theoret. Phys.* **72**, 471-479 (1984).

17. Jespersen, S., Metzler, R, Fogedby, H.C., Lévy flights in external force fields: Langevin and fractional Fokker-Planck equations, and their solutions, *Phys. Rev. E.* **59**, 2736 (1999).

18. Gorenflo, R., Mainardi, F., Moretti, D., Pagnini, G., & Paradisi, P., Discrete random walk models for space-time fractional diffusion, *Chem. Phys.* **284**(1/2), 521-541 (2002).

19. Saichev, A. & Zaslavsky, G. M., Fractional kinetic equations: solutions and applications, *Chaos* **7**(4), 753-764 (1997).

20. Feller, W., *An Introduction to Probability Theory and its Applications*, Vol. 2, 2nd Ed. (Wiely. 1971).

21. Martin, B. K., Wu, X. L. and Goldburg, W. I., Spectra of decaying turbulence in a soap film, 80(18), *Phys. Rev. Lett*., 3964-3967 (1998)

22. Mandelbrot, B., Intermittent turbulence in self-similar cascades: Divergence of high moments and dimension of the carrier, *J. Fluid Mech.* **62**, 331-358 (1974).

23. del-Castillo-Negrete, D., Carreras, B. A., Lynch, V. E. Non-diffusive transport in plasma turbulence: a fractional diffusion approach, *eprint arXiv:physics/0403039* (2004); Duran, I., Stockel, J., Hron, M., Horaÿcek, J., Jakubka, K., Kryska, L., Validity of self-organized criticality model for the CASTOR tokamak edge plasmas, *ECA Vol.* 24B *1693-1696 (2000)*.

24. Dubrulle, B. and Laval, J.-Ph., Truncated Levy laws and 2D turbulence, *Eur. Phys. J. B* **4**, 143-146 (1998).

25. Cardoso, O., Gluckmann, B., Parcollet, O. and Tabeling, P., Dispersion in a quasi-two-dimensional turbulent flow: an experimental study, *Phys. Fluids*, 8(1), 209-214 (1996).

26. Tsinober, A., Anomalous diffusion in geophysical and laboratory turbulence, *Nonlinear Processes in Geophysics*, 1, 80-94 (1994); Ragot, B. R., Transport of dust grains in turbulent molecular clouds, *Astrophysical Journal*, 498:757762, 1998.

27. Groisman, A. and Steinberg, V., Elastic turbulence in a polymer solution flow, *Nature*, **405**, 53-55 (2000).

28. Richardson, L. F., Atmospheric diffusion shown on a distance-neighbour graph, *Proc. R. Soc. London A*, 110, 709 (1926).

29. Chuvilgin, L. G. and Ptuskin, V. S., Anomalous diffusion of cosmic rays across the magnetic field, *A&A*, 279, 278 (1993).

30. Warhaft, Z., Passive scalars in turbulent flows, *Annu. Rev. Fluid Mech.* **32**, 203–240 (2000).